\documentclass[mathleft,fleqn,%
]{an}
\usepackage{graphicx}
\usepackage[varg]{txfonts}
\usepackage{bm}
\usepackage{natbib}
\bibpunct{(}{)}{;}{a}{}{,}
\graphicspath{{./fig/}{./png/}}
\newcommand{\uu}{\bm{u}}
\newcommand{\bra}[1]{\langle #1\rangle}
\newcommand{\Eq}[1]{Equation~(\ref{#1})}
\newcommand{\EQ}{\begin{equation}}
\newcommand{\EE}{\end{equation}}
\newcommand{\EQA}{\begin{eqnarray}}
\newcommand{\EEA}{\end{eqnarray}}

\newcommand{\pd}{\partial}

\newcommand{\urms}{u_{\rm rms}}

\newcommand{\Beq}{B_{\rm eq}}
\newcommand{\Ma}{{\rm Ma}}

\newcommand{\kef}{k_{\rm f}}

\newcommand{\Pm}{{\rm Pm}}
\newcommand{\Rm}{{\rm Rm}}
\newcommand{\Pra}{{\rm Pr}}
\newcommand{\Ra}{{\rm Ra}}

\newcommand{\Rey}{{\rm Re}}
\newcommand{\Rem}{{\rm Rm}}
\newcommand{\Remc}{{\rm Rm}_{\rm c}}

\newcommand{\nab}{\mbox{\boldmath $\nabla$} {}}

{}
\def\onethird{{\textstyle{1\over3}}}
\def\onehalf{{\textstyle{1\over2}}}

\newcommand{\Fig}[1]{Figure~\ref{#1}}

\begin{document}

% The following seven commands are intended for editorial usage and
% should be ignored by the author(s).
\Pagespan{1}{}% Document's page range. 
% If second parameter is left empty, the last page is computed
% automatically.
\Yearpublication{2014}%
\Yearsubmission{2014}%
\Month{0}%   
\Volume{999}%  
\Issue{0}% 
\DOI{asna.201400000}% 

%NORDITA-2018-011 

\title{Small-scale dynamos in simulations of stratified turbulent convection}

\author{Petri J. K\"apyl\"a\inst{1,2,3}\fnmsep\thanks{Corresponding author:
        {pkapyla@aip.de}}
\and  Maarit J. K\"apyl\"a\inst{3,2}
\and  Axel Brandenburg\inst{4,5,6,7}
}
\titlerunning{Small-scale dynamos in simulations of turbulent convection}
\authorrunning{P.\,J. K\"apyl\"a at al.}
\institute{Leibniz-Institut f\"ur Astrophysik Potsdam, 
  An der Sternwarte 16, D-11482 Potsdam, Germany
  \and ReSoLVE Centre of Excellence, Department of Computer Science,
  Aalto University, PO Box 15400, FI-00076 Aalto, Finland
  \and Max-Planck-Institut f\"ur Sonnensystemforschung,
  Justus-von-Liebig-Weg 3, D-37077 G\"ottingen, Germany
  \and NORDITA, KTH Royal Institute of Technology and Stockholm University,
  Roslagstullsbacken 23, SE-10691 Stockholm, Sweden
  \and Department of Astronomy, AlbaNova University Center,
  Stockholm University, SE-10691 Stockholm, Sweden
  \and JILA and Department of Astrophysical and Planetary Sciences,
  Box 440, University of Colorado, Boulder, CO 80303, USA
  \and Laboratory for Atmospheric and Space Physics,
  3665 Discovery Drive, Boulder, CO 80303, USA}

\received{26th Feb 2018}
%\accepted{XXXX}
%\publonline{XXXX}


\abstract{%
  Small-scale dynamo action is often held responsible for the
  generation of quiet-Sun magnetic fields. We aim to determine the
  excitation conditions and saturation level of small-scale dynamos in
  non-rotating turbulent convection at low magnetic Prandtl numbers.
  We use high resolution direct numerical simulations of weakly
  stratified turbulent convection.
  We find that the critical magnetic Reynolds number for dynamo
  excitation increases as the magnetic Prandtl number is decreased,
  which might suggest that small-scale dynamo action is not automatically evident in
  bodies with small magnetic Prandtl numbers as the Sun.
  As a function of the magnetic Reynolds number ($\Rem$), the
  growth rate of the dynamo is consistent with an $\Rem^{1/2}$ 
  scaling.
  No evidence for a logarithmic increase of the growth rate with $\Rem$
  is found.}

\maketitle

\section{Introduction}
Magnetic fields are ubiquitous in astrophysical systems. These fields
are in most cases thought to be generated by a dynamo
process, involving either turbulent fluid motions or
MHD-instabilities.
In dynamo theory \citep[e.g.][]{KR80,RH04,BS05,2012SSRv..169..123B} a
distinction is made between large-scale (LSD) and small-scale dynamos (SSD) where
the former produce fields whose length scale is greater than the scale
of fluid motions whereas in the latter the two are comparable.
Also an LSD can produce small-scale magnetic fields through 
tangling, 
and the decay of active regions will similarly cause magnetic energy to
cascade from larger to smaller scales, as is evidenced by the
presence of a Kolmogorov-type energy spectrum in their proximity
\citep{2014ApJ...784L..45Z,2016ApJ...819..146Z}.

Small-scale dynamos have been found in direct numerical simulations of
various types of flows provided that the magnetic Reynolds number
($\Rem$) exceeds a critical value ($\Remc$).
However, in many astrophysical conditions molecular kinematic
viscosity and magnetic diffusivity are vastly different implying that
their ratio, which is the magnetic Prandtl number ($\Pm$), is either
very small or very large. For example, in the Sun
$\Pm=10^{-3}\ldots10^{-6}$ \citep[e.g.][]{O03}. Numerical simulations
of forced turbulence
and other idealized flows indicate that $\Remc$ increases as $\Pm$ is
decreased \citep{PPP04,SCTMM04,SHBCMM05,SICMPY07,ISCMP07}. Theoretical
studies indicate a similar trend with an asymptotic value for $\Remc$
when $\Pm\rightarrow 0$ \citep{RK97}.
The work of \cite{ISCMP07} suggests that there is a value of $\Pm$ of
around $0.1$ where $\Remc$ is largest and that it decreases
again somewhat at even smaller values of $\Pm$.
In the nonlinear regime, however, no significant drop in the magnetic
energy is seen as $\Pm$ is decreased to and below $0.1$ \citep{Br11}.
More recently, \cite{SB14} found that the drop in the value of $\Remc$
may have been exaggerated by having used a forcing wavenumber that was
too close to the minimal wavenumber of the computational domain.

Simulations of turbulent convection have also been able to produce
SSDs
\citep[e.g.][]{MP89,NBJRRST92,BJNRST96,Cat99,PCS10,FB12,HRY15b}.
Such small-scale magnetic fields may explain the network of magnetic
fields in the Sun, which are independent of the solar cycle
\citep{2013A&A...555A..33B,St14,Re14}; see \cite{2017LRSP...14....4B} and
\cite{2017SSRv..210..275B} for reviews.
However, even the expected independence of the cycle does not go without
controversy \citep{2015AA...582A..95F,2016AA...585A..39U}.
In fact, \cite{Jin11} found evidence for an anticorrelation of
small-scale fields with the solar cycle.
This could potentially be explained by the interaction of the
SSD with superequipartition large-scale fields
from the global dynamo; see \cite{2016ApJ...816...28K}.
Small-scale magnetic fields may also play a role in heating the
solar corona; see \cite{ALA15} for recent work in that direction.

Small-scale dynamo-produced magnetic fields have been invoked
\citep{HRY15b,HRY16,2017ApJ...851...74B} to
explain the convective conundrum of the low levels of observed turbulent velocities
compared to contemporary simulations \citep[e.g.][]{2012PNAS..10911896G,MFRT12}.
Subsequent work of \cite{2018arXiv180100560K}, who studied cases
of large thermal Prandtl numbers conjectured to be due to the magnetic
suppression of thermal diffusion, does however cast some doubt on
this idea.

Returning to the problem of magnetic Prandtl numbers,
\cite{TS15} studied the case $\Pm\ge1$ from local solar surface
convection simulations and found that the SSD
ceases to exist already for $\Pm=1$.
However, this is mainly a shortcoming of low resolution.
Global and semi-global simulations of solar and stellar magnetism have
also recently reached parameter regimes where SSDs are
obtained \citep{HRY16,KKOWB16}. These models suggest that the vigorous
small-scale magnetism has profound repercussions for the LSD
and differential rotation. However, due to resolution
requirements the global simulations are limited to magnetic Prandtl
numbers of the order of unity or greater.

In the present paper, we therefore study high-resolution simulations
of convection-driven SSDs in the case of small values
of $\Pm$ by means of local models capturing more turbulent regimes.
This regime was already addressed in an early paper by \cite{Ca03},
but no details regarding the dependence of the growth rates and
saturation values of the magnetic field are available.

\section{The model}
\label{sec:model}

Our numerical model is the same as that of \cite{KKB10a} but without
imposed shear or rotation. We use a Cartesian domain with dimensions
$L_x=L_y=5d$ and $L_z=d$ with $0<z<d$, where $d$ is the depth of the
layer.

\subsection{Basic equations and boundary conditions}

We solve the set of equations of magnetohydrodynamics
\begin{eqnarray}
\frac{\pd \bm{A}}{\pd t} &=& \bm{U}\times\bm{B} - \eta \mu_0 \bm{J}, \\
\frac{D \ln \rho}{Dt} &=& -\bm\nabla\cdot\bm{U}, \\
\frac{D \bm{U}}{Dt} &=& {\bm g} -\frac{1}{\rho}\left[{\bm \nabla}p + \bm{J} \times {\bm B} + \bm{\nabla} \cdot (2 \nu \rho \mbox{\boldmath ${\sf S}$})\right],\label{equ:mom}\\
T\frac{D s}{Dt}\!&=&\!\frac{1}{\rho}\left(\mu_0\eta \bm{J}^2 - \bm\nabla\cdot {\bm F}^{\rm rad}\right) + 2 \nu \mbox{\boldmath ${\sf S}$}^2,
\end{eqnarray}
where $D/Dt=\pd/\pd t + \bm{U}\cdot\bm\nabla$ is the advective time
derivative, $\bm{A}$ is the magnetic vector potential, $\bm{B} =
\bm{\nabla} \times \bm{A}$ is the magnetic field, and $\bm{J} =
\mu_0^{-1} \bm{\nabla} \times \bm{B}$ is the current density, $\mu_0$
is the vacuum permeability, $\eta$ and $\nu$ are the magnetic
diffusivity and kinematic viscosity, respectively,
${\bm F}^{\rm rad}=-K\nab T$ is the radiative flux, $K$ is the
(constant) heat
conductivity, $\rho$ is the density, $\bm{U}$ is the velocity, $p$ is
the pressure and $s$ the specific entropy with $Ds=c_{\rm V} D \ln p -
c_{\rm P} D \ln \rho$, and
$\bm{g} = -g\hat{\bm{z}}$ is the gravitational acceleration. The fluid
obeys an ideal gas law $p=\rho e (\gamma-1)$, where $p$ and $e$ are
the pressure and internal energy, respectively, and $\gamma = c_{\rm
  P}/c_{\rm V} = 5/3$ is the ratio of specific heats at constant
pressure and volume, respectively. The specific internal energy per
unit mass is related to the temperature via $e=c_{\rm V} T$. The rate
of strain tensor $\mbox{\boldmath ${\sf S}$}$ is given by
\begin{equation}
{\sf S}_{ij} = \onehalf (U_{i,j}+U_{j,i}) - \onethird \delta_{ij} \bm\nabla\cdot\bm{U}.
\end{equation}

In order to exclude complications due to overshooting and compressibility,
we employ a weak stratification: the density difference between the
top and bottom of the domain is twenty per cent and the average
Mach number, $\Ma=\urms/\sqrt{dg}$, is always less than 0.1. The stratification in the
associated hydrostatic initial state can be described by a polytrope
with index $m=1$.
The stratification is controlled by the normalized pressure scale
height at the surface
\begin{eqnarray}
\xi_0 = \frac{(c_{\rm P}-c_{\rm V}) T_1}{gd},
\end{eqnarray}
where $T_1$ is the temperature at the surface ($z=d$). In our current
simulations we use $\xi_0=2.15$.

The horizontal boundary conditions are periodic. We keep the temperature fixed
at the top and bottom boundaries. For the velocity we apply
impenetrable, stress-free conditions according to
\begin{eqnarray}
\pd_zU_x = \pd_z U_y = U_z=0.
\end{eqnarray}
For the magnetic field we use vertical field conditions
\begin{eqnarray}
B_x = B_y=0.
\end{eqnarray}

\begin{figure*}[t]
\centering
\includegraphics[width=\textwidth]{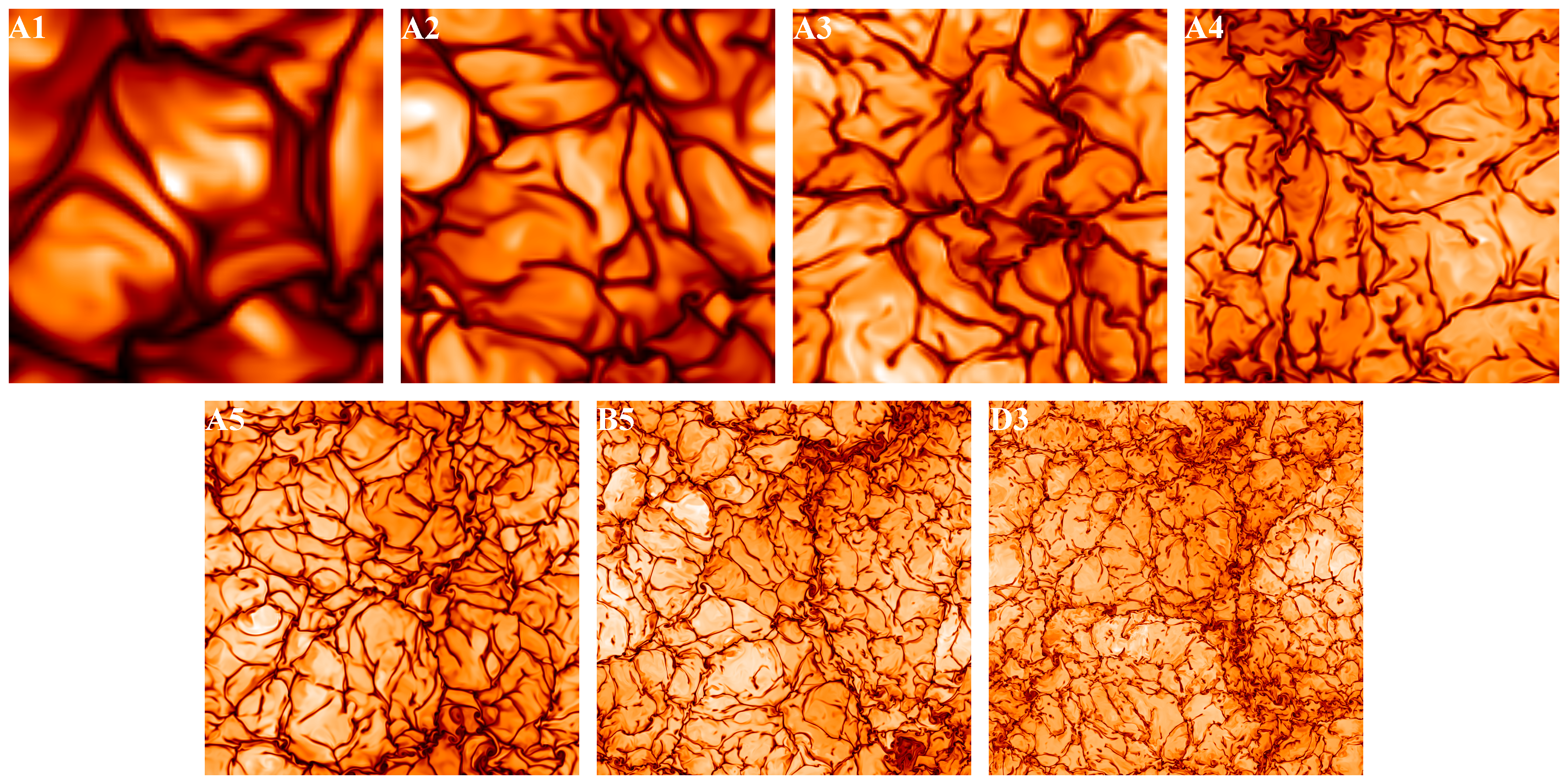}
\caption{Upper row: specific entropy $s/c_{\rm P}$ near the surface
  $z/d=0.98$ for $\Rey={\rm Pe}=$ 23, 54, 101, and 193. Lower row:
  $\Rey={\rm Pe}=$ 354, 666, and 1057.}
\label{fig:ss_slices}
\end{figure*}

\subsection{Units, nondimensional quantities, and parameters}

The units of
length, time, velocity, density, specific entropy, and magnetic field are then
\begin{eqnarray}
&& [x] = d\;,\;\; [t] = \sqrt{d/g}\;,\;\; [U]=\sqrt{dg}\;,\;\; \nonumber \\ && [\rho]=\rho_0\;,\;\; [s]=c_{\rm P}\;,\;\; [B]=\sqrt{dg\rho_0\mu_0}\;,
\end{eqnarray}
where $\rho_0$ is the density of the initial state at $z_{\rm m}=\onehalf d$.
The simulations are controlled by the following dimensionless
parameters: thermal and magnetic diffusion in comparison to viscosity
are measured by the Prandtl numbers
\begin{eqnarray}
\Pr=\frac{\nu}{\chi_0}, \quad \Pm=\frac{\nu}{\eta},
\end{eqnarray}
where $\chi_0=K/(c_{\rm P} \rho_0)$ is the reference value of the
thermal diffusion coefficient, measured in the middle of the layer,
$z_{\rm m}$, in the non-convecting initial state. The efficiency of
convection is measured by the Rayleigh number
\begin{eqnarray}
\Ra=\frac{g d^4}{\nu \chi_0}\left(- \frac{1}{c_{\rm P}}\frac{{\rm d}s}{{\rm d}z} \right)_{z_{\rm m}},
\end{eqnarray}
again determined from the initial non-convecting state at $z_{\rm m}$. The
entropy gradient can be presented as
\begin{eqnarray}
\left(- \frac{1}{c_{\rm P}}\frac{{\rm d}s}{{\rm d}z} \right)_{z_{\rm m}}=\frac{\nabla-\nabla_{\rm ad}}{H_{\rm P}},
\end{eqnarray}
where $\nabla=(\pd \ln T/\pd \ln p)_{z_{\rm m}}$ and
$\nabla_{\rm ad}=1-1/\gamma$ are the actual and adiabatic
double-logarithmic temperature gradients
and $H_{\rm P}$ is the pressure scale height at $z=z_{\rm m}$.

The effects of viscosity and magnetic diffusion are quantified
respectively by the fluid and magnetic Reynolds numbers
\begin{eqnarray}
\Rey=\frac{\urms}{\nu \kef}, \quad \Rem=\frac{\urms}{\eta \kef}=\Pm\,\Rey,
\end{eqnarray}
where $\urms$ is the root mean square (rms) value of the velocity, and
$\kef=2\pi/d$ is the wavenumber corresponding to the depth of the
layer.
Furthermore, we define the P$\acute{\rm e}$clet number as 
\begin{eqnarray}
{\rm Pe}=\frac{\urms}{\chi_0 \kef} = \Pra\,\Rey.
\end{eqnarray}
Except for the simulations of Section~\ref{SatLevel}, where $\Pra=\Pm$ is varied,
we use in all other simulations $\Pra=1$ and thus ${\rm Pe}=\Rey$.

Error estimates are obtained by dividing the time series into
three equally long parts. The largest deviation of the average for
each of the three parts from that over the full time series is taken
to represent the error.

The simulations were performed using the {\sc Pencil Code}%
\footnote{https://pencil-code.github.com/}, which uses
sixth-order explicit finite differences in space and a third-order
accurate time stepping method.
We use resolutions ranging from $64^3$ to $1024^3$.

\begin{table}[t!]
\centering
\caption[]{Summary of weak field runs for $\Pr=1$.}
% The runs are found in the folder: \texttt{pencil-code/petri/convection/sscaledyn}
      \label{tab:runs}
     $$
         \begin{array}{p{0.07\linewidth}cccccrrcc}
           \hline
           \noalign{\smallskip}
Run & \Pm & \Ra [10^6] & \Ma & \Rey & \Rem & \tilde\lambda [10^{-4}] & \delta\tilde\lambda [10^{-4}] & $grid$ \\ \hline 
A1 &   1   & 0.17 & 0.073 &   23 &  23 &  -52 & 47 &   64^3 \\ % 64a1
A2 &   1   & 1.0  & 0.068 &   54 &  54 &   62 & 18 &  128^3 \\ % 128a3
A3 &   1   & 4.2  & 0.064 &  101 & 101 &  162 & 20 &  128^3 \\ % 128a1
A4 &   1   & 17 & 0.061 &  193 & 193 &  273 &  7 &  256^3 \\ % 256a1
A5 &   1   & 67  & 0.056 &  354 & 354 &  453 &  2 &  512^3 \\ % 512a1
           \hline
B1 &  0.5  & 1.0  & 0.069 &   54 &  27 & -128 & 24 &  128^3 \\ % 128a4
B2 &  0.5  & 4.2 & 0.064 &  102 &  51 &  -27 & 12 &  128^3 \\ % 128a2
B3 &  0.5  & 17 & 0.060 &  191 &  95 &   44 & 11 &  256^3 \\ % 256a2
B4 &  0.5  & 67 & 0.056 &  360 & 180 &  155 & 18 &  512^3 \\ % 512a2
B5 &  0.5  & 267 & 0.052 &  666 & 333 &  357 & 17 & 1024^3 \\ % 1024a3
           \hline
C1 & 0.25  & 17 & 0.060 &  190 &  47 & -144 & 22 &  256^3 \\ % 256a3
C2 & 0.25  & 67 & 0.056 &  358 &  90 &  -35 & 28 &  512^3 \\ % 512a3
           \hline
D1 &  0.1  & 67 & 0.057 &  360 &  36 & -237 & 56 &  512^3 \\ % 512a4
D2 &  0.1  & 267 & 0.052 &  664 &  65 & -139 & 77 & 1024^3 \\ % 1024a1
D3 &  0.1  & 740 & 0.050 & 1057 & 106 &  -10 & 94 & 1024^3 \\ % 1024a2
           \hline
         \end{array}
     $$
\end{table}

\begin{figure}[t]
\centering
\includegraphics[width=\columnwidth]{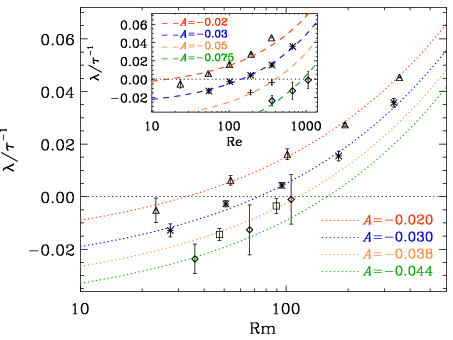}
\caption{Growth rate $\lambda$ of the rms magnetic field normalized by
  the inverse convective turnover time $\tau^{-1}=\urms \kef$ as a
  function of the magnetic Reynolds number $\Rem$. The different
  symbols denote runs with $\Pm=1$ (triangles), $\Pm=0.5$ (stars),
  $\Pm=0.25$ (squares), and $\Pm=0.1$ (diamonds). The horizontal dotted
  line denotes marginal stability. The red, blue, orange, and green dotted lines are
  curves proportional to $\Rem^{1/2}$; see \Eq{eq:tildeg} for different
  values of $A$ and $B=3.5\cdot10^{-3}$ is fixed. The inset shows the
  normalized growth rates for the same data as functions of $\Rey$. The
  dashed lines are proportional to $\Rey^{1/2}$ according to a relation
  analogous to \Eq{eq:tildeg} with $B=2.5\cdot10^{-3}$ and values of
  $A$ indicated in the legend.}
\label{fig:pdata}
\end{figure}

\section{Results}

\subsection{Description of the runs}

We perform four sets of runs where we keep the magnetic Prandtl
number fixed and vary $\Rey$ and $\Rem$; see Table~\ref{tab:runs}. The
lower resolution ($64^3$, $128^3$, and $256^3$) runs were started from
a non-convecting state described in the previous section, whereas runs
at $512^3$ and $1024^3$ were remeshed from saturated snapshots at
lower resolutions; see \Fig{fig:ss_slices} for visualizations of
specific entropy near the surface of the domain.
After the convection has reached a statistically
saturated state we introduce a weak random magnetic field of the order
of $10^{-6}\Beq$, where $\Beq$ is the equipartition field strength
with $\Beq^2=\bra{\mu_0\rho\uu^2}$.
We refer to these runs as weak field models and perform the data
analysis in regimes where the magnetic fields remains dynamically
unimportant.
After an initial transient, the growth rate of the
total rms magnetic field is measured as
\begin{eqnarray}
\lambda=\langle d \ln B_{\rm rms}/dt \rangle_t,
\end{eqnarray}
where $\langle \cdots \rangle_t$ denotes time averaging. In the runs
where the dynamo is clearly above or below critical, a short time
series (few tens of turnover times) is sufficient to measure a
statistically significant value of $\lambda$. The runs near the excitation
threshold need to be run significantly longer (hundreds of turnover
times). For the highest resolution runs at low $\Pm$ this is not
feasible due to the computational cost and thus the error bars for
these runs are typically significantly larger than in the low resolution
or $\Pm=1$ runs.

\subsection{Growth rate in the kinematic regime}

Figure~\ref{fig:pdata} shows the growth rate of the magnetic field as a
function of $\Rem$ for the four magnetic Prandtl numbers explored in
the current study.
For reference, we plot curves of the form
\begin{eqnarray}
\gamma/\tau^{-1} \equiv \tilde\gamma = A + B\,\Rm^{1/2},
\label{eq:tildeg}
\end{eqnarray}
where the
value of the constant $A$ changes as the magnetic Prandtl number is
changed. Furthermore, $B=3.5\cdot10^{-3}$ for all values of $\Pm$.
The parameter $A$ is negative, so the solutions will always decay
for small values of $\Rm$, but they increase with increasing
values of $\Pm$; see \Fig{fig:pdata}.

We find that the normalized growth rate for a given $\Rm$ decreases as
$\Pm$ is decreased. Surprisingly, $\tilde\lambda$ appears to follow a
$\Rm^{1/2}$ trend for each value of $\Pm$ -- even in the cases when an
SSD is not excited. Such a dependence is predicted by theory for high
$\Rm$, i.e.\ far away from excitation \citep{KR12}. However, given the
relatively large error bars, the $\Rm^{1/2}$ scaling near the
excitation threshold can at this point only be suggestive and far from
definitive. Indeed, analytic theory yields a different scaling in this
regime \citep{KR12}.

In the low-$\Pm$ regime, the growth rate of the magnetic field due to
the SSD is expected to scale with the 1/2 power of the fluid Reynolds
number. We find that our simulation data is consistent with this
for values of $\Pm$ of 0.5 and smaller; see the inset to
Figure~\ref{fig:pdata}.

\subsection{Dependence on the box size}

\begin{figure}[t]
\centering
\includegraphics[width=\columnwidth]{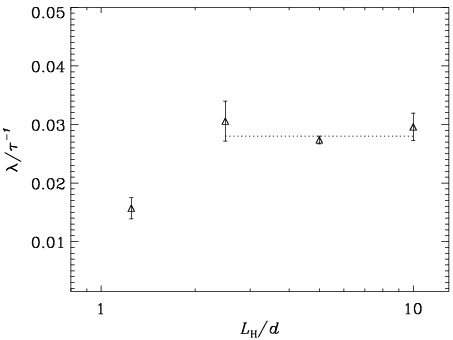}
\caption{Growth rate $\lambda$ of the rms magnetic field normalized by
  the inverse convective turnover time $\tau^{-1}=\urms \kef$ as a
  function of the horizontal box size for $\Pm=1$.}
\label{fig:pdata_size}
\end{figure}

The dependence of the growth rate of the convection-driven SSD
on the horizontal size of the domain and the presence of
mesogranulation has been discussed in a recent paper by
\cite{BFPW12}.
To address this issue, we show in Figure~\ref{fig:pdata_size} the growth rate of
the magnetic field as a function of the horizontal box size for
magnetic Prandtl number unity. Deviations from the constant trend are
found for $L_{\rm H}/d=1.25$. For $L_{\rm H}/d=0.5$, no dynamo
action is found. Our standard box size of $L_{\rm H}/d=5$ is thus
adequate and does not seem to suffer from the issues raised by
\cite{BFPW12}.
This is in spite of the fact that the flow is dominated by a single
convection cell filling the whole domain, which is clear even by
visual inspection of Figure~\ref{fig:ss_slices}.

\begin{figure}[t]
\centering
\includegraphics[width=\columnwidth]{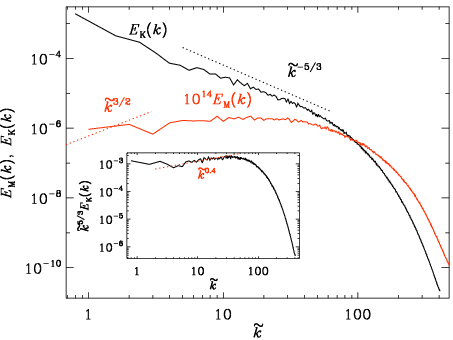}
\caption{Power spectra of velocity and magnetic field as functions of
  $\tilde{k}=k/k_1$ near the top of the domain from Run~B5. $E_{\rm
    M}$ has been multiplied by $10^{14}$ for visualization
  purposes. The dotted line show a $k^{-5/3}$ scaling for
  reference. The inset shows the velocity power spectrum compensated
  by $k^{5/3}$.}
\label{plot_spectra_1024a3}
\end{figure}

\subsection{Energy spectra}

In \Fig{plot_spectra_1024a3} we show the kinetic and magnetic energy spectra,
$E_{\rm K}$ and $E_{\rm M}$, respectively, for Run~B5 during the kinematic
phase of the dynamo for $\Pm=0.25$ and $\Rm=654$.
The kinetic energy spectrum shows a clear $k^{-5/3}$ spectrum along with
a slightly shallower slope near the dissipative cutoff.
This is the bottleneck effect \citep{Fal94}, which has been held
responsible for causing the increase of $\Remc$ near $\Pm=0.1$, because
then the peak of the magnetic energy lies fully within the inertial
range of the kinetic energy spectrum \citep{BC04}.
The magnetic energy spectrum, on the other hand, is significantly
shallower than the $k^{3/2}$ spectrum expected from the work of
\cite{Kaz68}, which has been confirmed in many in several numerical
simulations of kinematic dynamo action in forced turbulence
\citep{HBD04,SCTMM04} and supernova-driven turbulence
\citep{2004ApJ...617..339B}.
The spectra shown in \Fig{plot_spectra_1024a3} were taken from a run
where the magnetic field has grown only by a factor of a few and it is
possible that the 3/2 scaling has not had enough time to develop
yet. However, the flow exhibits a long-lived large-scale
component, manifested by the peak at $\tilde{k}=1$, which is not
present in simpler forced turbulence simulations. Such flows may
contribute to the relatively high magnetic power at large scales.
In that case, the lack of a $k^{3/2}$ spectrum in the kinematic
regime would have a physical origin.
These aspects will be explored further elsewhere.

\begin{table}[t!]
\centering
\caption[]{Summary of strong field runs.}
% The runs are found in the folder: \texttt{pencil-code/petri/convection/sscaledyn}
      \label{tab:runs2}
     $$
         \begin{array}{p{0.07\linewidth}ccccrrcc}
           \hline
           \noalign{\smallskip}
Run & \Pm & \Pr & \Ra [10^6] & \Rey & \Rem & \Ma & \tilde{B}_{\rm rms} & $grid$ \\ \hline 
S1 &    1   &   1  &  17 &  169 &  169 & 0.053 & 0.0130 &  256^3 \\ % 256a1_cont
S2 &   0.5  &  0.5 &  33 &  361 &  180 & 0.057 & 0.0126 &  256^3 \\ % 256b1
S3 &   0.25 & 0.25 &  67 &  760 &  190 & 0.060 & 0.0110 &  256^3 \\ % 256b2
S4 &   0.1  &  0.1 & 167 & 2118 &  212 & 0.067 &    -   &  512^3 \\ % 512b1
           \hline
         \end{array}
     $$
\end{table}

\subsection{Saturation level}
\label{SatLevel}

Another set of simulations was made to study the saturation level of
the magnetic fields produced by the SSD; see Table~\ref{tab:runs2}.
We refer to these runs as strong field models. These runs were either
run to full saturation from the initial conditions described in
Section~\ref{sec:model} (Run~S1) or continued from a saturated snapshot
of an earlier run (S2, S3, or S4). At each step, the kinematic viscosity
is lowered to decrease $\Pm$ with the aim of avoiding the long
kinematic stage of the dynamo. Another possible advantage of this procedure is that the
SSD has been shown to operate in the nonlinear regime at an $\Rm$
value that would be subcritical in the kinematic case
\citep{Br11}. While this procedure works for Runs~S2 and S3, in Run~S4
with $\Pm=0.1$ where $\Rm=212$ and $\Rey=2118$ the magnetic field is
not sustained, however.

\begin{figure}[t]
\centering
\includegraphics[width=\columnwidth]{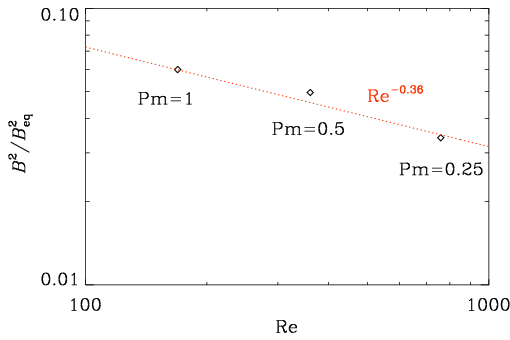}
\caption{Saturation field strength for Runs~S1 to S3 with $\Pm=1\ldots 0.25$
  ($\Rem=169\ldots190$) as a function of the fluid Reynolds number.}
\label{fig:pdata_sat}
\end{figure}

Figure~\ref{fig:pdata_sat} shows the saturation field strength for
Runs~S1 to S3 with
$\Pm=0.25\ldots 1$. Here the magnetic Reynolds number varies from 169
to 190 due to increasing $\urms$ when $\Pm$ decreases.
Contrary to the results for forced turbulence, where the rms magnetic
field was found to decrease by no more than a factor of two as $\Pm$
was decreased from unity to $0.01$ \citep{Br11}, we seem to find here
a somewhat stronger dependence of the saturation field strength on the
value of $\Pm$ and thus on $\Rey$.
The current results suggest a scaling with $\Rey$ with a power that is
close to $-1/3$.
However, one has to realize that for the run with the largest value of $\Rey$
and $\Pm=0.25$, we used a resolution of $256^3$ which may be too low to
resolve the flow at $\Rey=760$.

\section{Conclusions}

Our work has confirmed that in turbulent convection at low values of
$\Pm$, the value of $\Remc$ increases with decreasing $\Pm$.
This effect may well be connected with the bottleneck effect seen in
the kinetic energy spectrum.
The saturated field strength, however, is found to show a somewhat
stronger dependence on $\Pm$ than in the case of forced turbulence.

Both for small values of $\Pm$ and for $\Pm$ of unity, we find that
the kinematic growth rate increases proportional to $\Rm^{1/2}$.
In particular, there is no evidence for a logarithmic dependence.
A similar dependence on $\Rm$ has previously been seen in forced
turbulence; see \cite{HBD04}, for example.

Interestingly, however, in the kinematic regime, the magnetic energy
spectrum is significantly shallower than the $k^{3/2}$ spectrum expected
for an SSD \citep{Kaz68}.
This is also quite different from the case of forced turbulence, where
a clear $k^{3/2}$ spectrum is found during the kinematic growth phase.
In other words, the current kinematic convection-driven dynamo show a tendency
of producing larger-scale magnetic fields than in forced turbulence.
This is possibly caused by a persistent large-scale velocity pattern
which is a robust feature in the current simulations.

\acknowledgements
  The computations were performed on the facilities hosted by CSC --
  IT Center for Science Ltd. in Espoo, Finland, who are administered
  by the Finnish Ministry of Education. We also acknowledge the
  allocation of computing resources provided by the Swedish National
  Allocations Committee at the Center for Parallel Computers at the
  Royal Institute of Technology in Stockholm. This work was supported in part
 by the Academy of Finland ReSoLVE Centre of Excellence (grant
No.\ 272157; MJK \& PJK),
  the NSF Astronomy and Astrophysics Grants Program (grant 1615100), and
  the University of Colorado through its support of the George Ellery Hale
  visiting faculty appointment.

\bibliographystyle{an}
\bibliography{../bibtex/bib}

\end{document}